\documentclass{nature}
\usepackage{graphicx}
\usepackage{amssymb}
\usepackage{amsmath}
\usepackage{empheq}
\usepackage{mathrsfs}
\usepackage{stmaryrd}
\usepackage[cyr]{aeguill}
\usepackage[english]{babel}
\usepackage[linktocpage]{hyperref}
\usepackage{tocvsec2}
\usepackage{color}
\usepackage{version}
\usepackage{textcomp}
\usepackage{longtable}
\usepackage{textcomp}
\usepackage{gensymb}
\usepackage{wasysym}
\usepackage{float}

\title{Hot Hydride Superconductivity above 550 K }

\author{A. D. Grockowiak*$^{1}$, M. Ahart$^2$, T. Helm$^{3,4}$, W.A. Coniglio$^{1}$, R. Kumar$^2$, M. Somayazulu$^5$, Y. Meng$^5$, M. Oliff$^{1}$, V. Williams$^{1}$, N.W. Ashcroft$^6$, R. J. Hemley$^{2,7}$, \& S. W. Tozer* $^1$}

\begin{document}

\maketitle
\\$^*$corresponding authors
\begin{affiliations}
 \item National High Magnetic Field Laboratory, Florida State University, Tallahassee, Florida, 32310, USA
 \item  Department of Physics, University of Illinois Chicago, Illinois 60607, USA
 \item  Dresden High Magnetic Field Laboratory (HLD-EMFL), Helmholtz-Zentrum Dresden-Rossendorf, 01328 Dresden, Germany
 \item Max-Planck-Institute for Chemical Physics of Solids, Noethnitzer Strasse 40, D-01187 Dresden, Germany
 \item  HPCAT, X-ray Science Division, Argonne National Laboratory, Lemont IL 60439, USA
 \item  Laboratory of Atomic and Solid State Physics, Cornell University, Ithaca, New York 14853, USA
 \item Department of Chemistry, University of Illinois Chicago, Illinois 60607, USA
\end{affiliations}

\begin{abstract}
The search for room temperature superconductivity has accelerated dramatically in the last few years driven largely by theoretical predictions that first indicated alloying dense hydrogen with other elements could produce conventional phonon-mediated superconductivity at very high temperatures  and at accessible pressures, and more recently, with the success of structure search methods that have identified specific candidates and pressure-temperature (P-T) conditions for synthesis. These theoretical advances have  prompted improvements in experimental techniques to test these predictions. As a result, experimental studies of simple binary hydrides under pressure have yielded  high critical superconducting transition temperatures (T$_{c}$), of 260 K in LaH$_{10}$, close to the commonly accepted threshold for room temperature, 293 K, at pressures near 180 GPa. We successfully synthesized a metallic La-based superhydride from La metal and  ammonia borane, NH$_{3}$BH$_{3}$, and find a  multi-step transition with a T$_{c}$ of 294 K for the highest onset.  When subjected to subsequent thermal excursions to higher temperatures that promoted a chemical reaction to what we believe is a ternary or higher order system, the transition temperature was driven to higher temperatures. Although the reaction does not appear to be complete, the onset temperature  was pushed from 294 K to 556 K before the experiments had to be terminated. The results provide evidence for hot superconductivity well above room temperature, in line with recent predictions for a higher order hydride under pressure.
\end{abstract}

\section*{Introduction}
Ever since Kamerlingh Onnes'  1911 discovery of superconductivity in mercury \cite{Onnes1911}, scientists have searched for materials with higher transition temperatures, initially in the elements, and then progressing to more complex systems. Although largely curiosity driven, each discovery of a material with higher critical field and temperature addresses clear technological needs. Fully fifty-five elements are now known to be superconducting at ambient or high pressure \cite{Schilling2007},  with hydrogen being a notable exception. In 1968 Ashcroft \cite{Ashcroft1968}   proposed that atomic metallic hydrogen  at sufficiently high density could be a very high T$_{c}$ Bardeen-Cooper-Schrieffer (BCS) superconductor \cite{Bardeen1957}. In view of the high pressures required to reach this proposed state of hydrogen, Carlsson and Ashcroft \cite{Carlsson1983} later suggested alternative routes to effectively produce superconducting atomic hydrogen, including the incorporation of other elements in the structure. This prediction prompted the experimental search for such compounds and alloys\cite{Eremets2003}.  Ashcroft\cite{Ashcroft2004} later extended and recast the above considerations in terms of ‘chemical pre-compression’, a proposal in which H$_{2}$ molecules in dense structures might be expected to dissociate at pressures well below those required for pure hydrogen.  Advances in crystal structure prediction methods then began to provide experimentalists specific targets for higher T$_{c}$ BCS superconductivity \cite{Zurek_2015}$^{,}$\cite{Wang_2014}$^{,}$\cite{Zhang2017}$^{,}$\cite{Needs2016}$^{,}$\cite{Oganov2019}$^{,}$\cite{Semenok2018} while at the same time theoretical studies have focused on understanding the superconducting mechanism \cite{Tanaka2017}$^{,}$\cite{Errea2020}$^{,}$\cite{Heil2019}$^{,}$\cite{Quan2019}$^{,}$\cite{Liu2019}$^{,}$\cite{Wang2019}$^{,}$\cite{FloresLivas2019}. Notably, this effort led to the prediction  \cite{Li2014}$^{,}$\cite{Duan2014} and, independently, to the experimental discovery \cite{Drozdov2015} of superconductivity in the H-S system with a T$_{c}$ of 203 K, including an isotope study that pointed to BCS behavior.  Subsequent simulations \cite{Liu2017}$^{,}$\cite{Peng2017}  guided the discovery of new hydrides with T$_{c}$ approaching room temperature, the highest being LaH$_{10}$ with a critical temperature of at least 260 K, \cite{Hemley2019}$^{,}$ \cite{Somayazulu2019}$^{,}$\cite{Drozdov2019}  and somewhat lower values of 227-243 K reported for related YH$_{6}$/YH$_{9}$ phases \cite{Troyan2019}$^{,}$\cite{Kong2019}. \\

What had once been proposed as a goal in and of itself, superconductivity at room temperature now appears to be the stepping off point for still higher T$_{c}$\textquotesingle s. Much like when the observation of superconductivity in cuprates led to a flurry of discoveries and engineering applications that pushed the critical temperatures in these materials from 40 to 164 K over the course of a decade \cite{Bednorz1986}$^{,}$\cite{Gao1994},  hydrogen-based materials hold out great promise. Current calculations predict that crystalline atomic metallic hydrogen \cite{Ashcroft1968}$^{,}$\cite{Wigner1935}  has a T$_{c}$ near room temperature  at 500 GPa, which increases to 420 K above 3 TPa \cite{McMahon2012}. Other calculations predict that hydrogen could be a superconducting superfluid at comparable conditions \cite{Babaev2004}.  Emerging systematics for the binary hydrides indicate maximum T$_{c}$ in the vicinity of room temperatures at megabar pressures \cite{FloresLivas2019}, and there is now a focus on the possibility of higher critical temperatures in more chemically complex hydride systems. Indeed, recent theoretical studies predict a T$_{c}$ of 473 K in Li$_{2}$MgH$_{16}$ near 250 GPa \cite{Sun2019}. \\

The present work was motivated by a desire to extend the P-T-H phase diagram of  previous measurements of the magnetic field dependence of the lanthanum-based superhydrides to fields approaching 100 T \cite{Drozdov2019}$^{,}$\cite{Liu2017}$^{,}$\cite{Geballe2018}. We also sought to examine lower pressure phases such as LaH$_{6}$ which is more experimentally accessible.  In addition we hoped to better understand the variable T$_{c}$ in experiments in which LaH$_{10}$ superhydride was synthesized using ammonia borane (NH$_{3}$BH$_{3}$) as the hydrogen source\cite{Somayazulu2019}.  For this purpose, we developed metallic DACs designed for superhydride studies in DC magnetic fields that are also small enough to fit into pulsed magnets.  Made from the highly resistive non-magnetic superalloy NiCrAl [Pascalloy, Tevonics] to limit the eddy-current heating, these DACs can be used to access temperatures down to at least 30 K in pulsed fields.\\

 We coupled this design with robust Pt electrodes created by Focused Ion Beam (FIB) techniques to withstand the extreme P-T conditions required to synthesize the superhydrides \cite{Helm2019}.  Attempts to characterize in situ by x-ray diffraction were unsuccessful due to an insufficient downstream opening in the DAC. We find that the La-based superhydride initially synthesized by laser-heating beginning at 160 GPa had a T$_{c}$ of 294 K. Most remarkably, subjecting the sample to subsequent thermal cycling shifted the T$_{c}$ to higher temperatures in a fortuitous  progression that reached well above 500 K. \\

\section*{Results}
\begin{figure}[]
\begin{minipage}[b]{0.45\linewidth}
\includegraphics[scale=0.45]{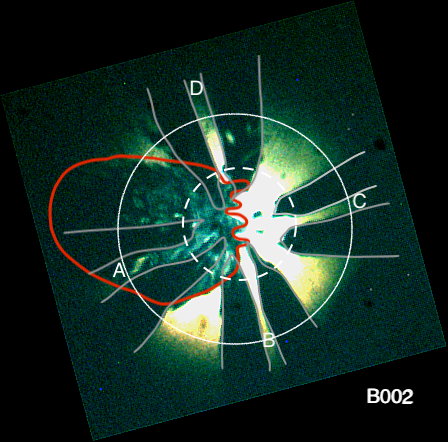}
\end{minipage}
\begin{minipage}[b]{0.45\linewidth}
\includegraphics[scale=0.65]{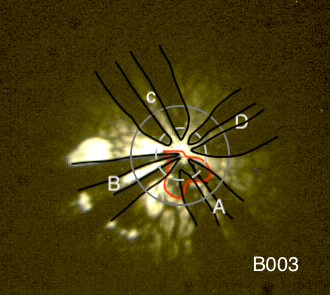}
\end{minipage}
\caption{\it Image of the sample and electrodes in cells B002 (left) and B003 (right) taken through the DAC with back-illumination. The letters indicate electrode pairs, outlined in grey. The red outline indicates the smeared initial piece of La.  The outer circle indicates the approximate boundary of the metal gasket and the cBN; the inner dashed circle indicates the perimeter of the 72 µm culet. After compression, both cells presented electrical shorts between the gasket and the electrodes. The initial amount of La inserted into each cell was also different: a thicker flake was introduced into B002, which upon compression extruded outside the culet region where it smeared across some electrodes as shown. Six and four electrodes are connected to the sample in B002 and B003, respectively. The FIBed electrodes run down the pavilion of the anvils where they are connected to copper twisted pairs using Epo-Tek H20E epoxy. }
	\label{figure1.fig}
\end{figure}

\begin{figure}[]
\begin{minipage}[b]{0.45\linewidth}
\includegraphics[scale=0.25]{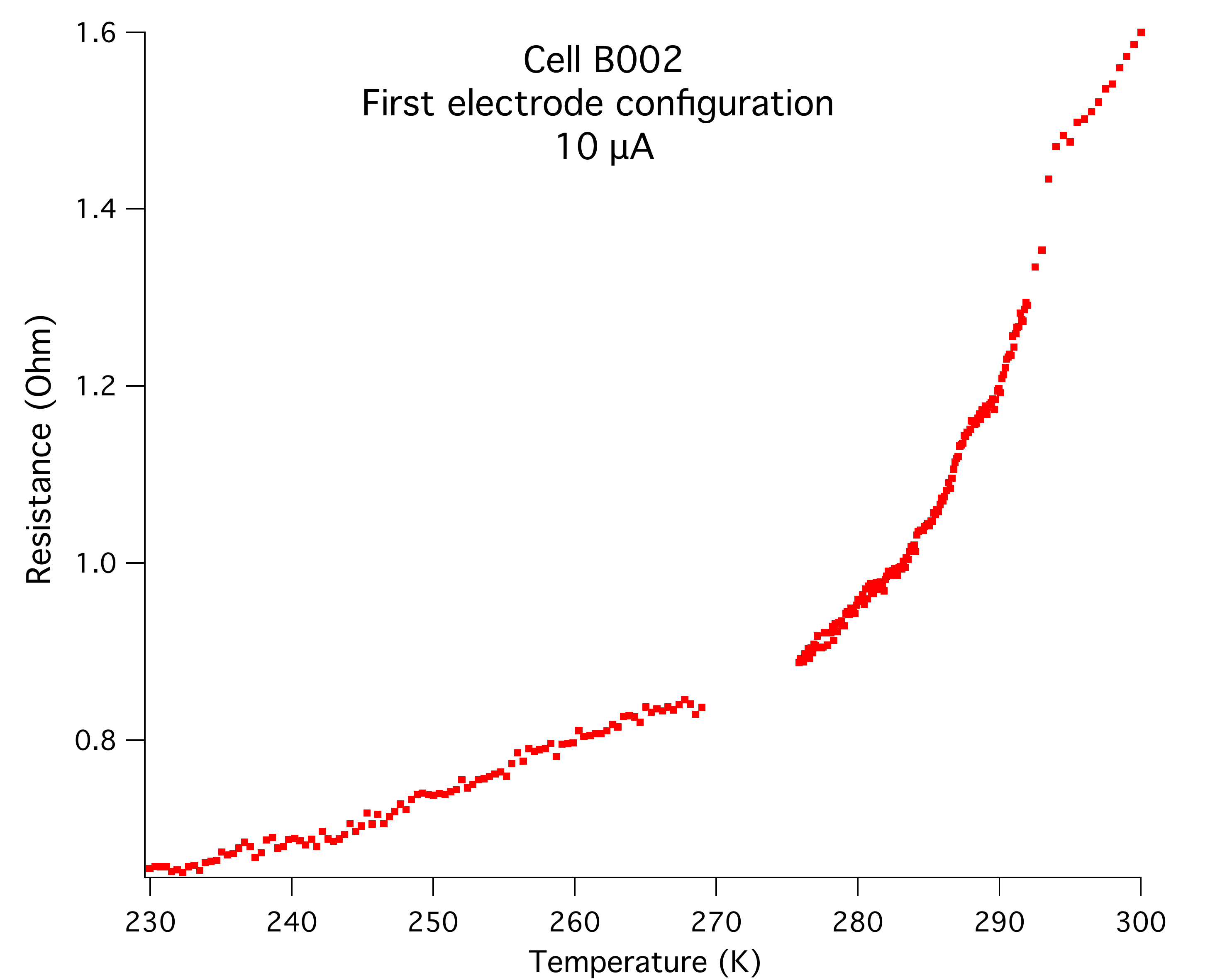}
\end{minipage}
\begin{minipage}[t]{0.5\linewidth}
\includegraphics[scale=0.25]{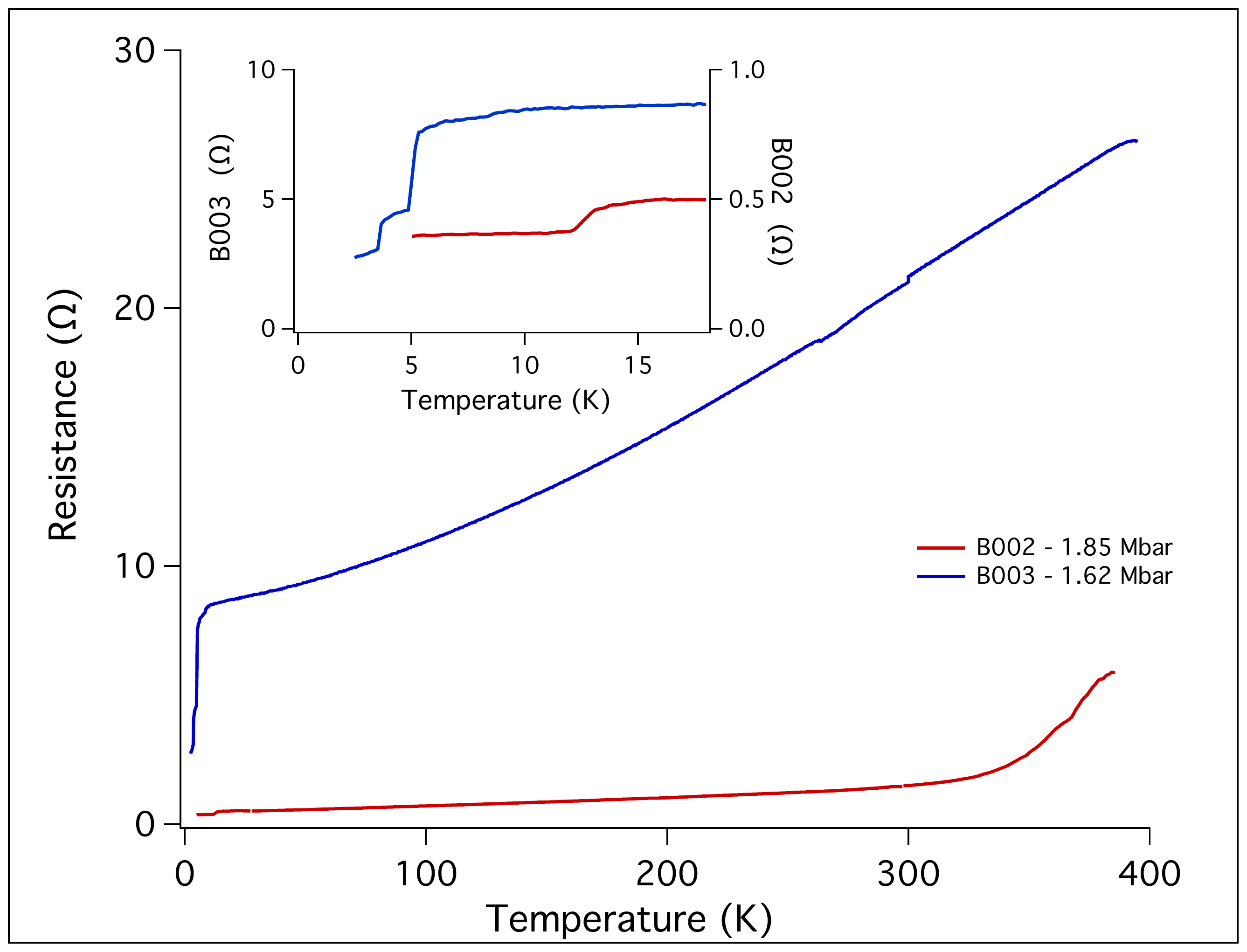}
\end{minipage}
\caption{\it  4-probe resistance measurements for the samples in cells B002 and B003.  Left: Results obtained using a first set of electrodes for B002. The current used was 10 $\mu$A, and the temperature cool down rate was 1 K/min. There is a clear superconducting transition at  294 K which is at the upper theoretical limit for the binary LaH$_{10}$ \cite{Peng2017}$^{,}$\cite{Liu2017}. (see SI) Data between 268 K and 277 K was not collected due to an error. Right: Comparison of B002 and B003.  Both show the superconducting transition due to unreacted La or a lower stoichiometry hydride at temperatures below 5 K and a superconducting transition at 12.5 K in B002 and 9 K in B003 that we attribute to a platinum hydride \cite{Scheler_2011}. The transition in B003 is at a lower temperature, consistent with B003 being at a lower pressure. It is also not as fully developed, which we attribute to the small number of laser pulses that B003 was subjected to.  B002 has the additional transition at temperatures greater than 365 K. Both B002 and B003 have a non-zero resistance down to 1.9 K. The resistance of B003 is almost a factor of 10 higher than B002, which is attributed to the thinness of the B003 sample.}
	\label{figure2.fig}
\end{figure}

We loaded two diamond-anvil cells (DACs) with pieces of 99$\%$ La and ammonia borane (AB, NH$_{3}$BH$_{3}$), the hydrogen source and pressure medium (see Methods and SI). Both cells were initially as identical as experimentally possible, with the same FIB-patterned electrodes on one anvil and an MP35N gasket with cBN insert and AB on the other. The first cell (B002) was loaded initially at 160 GPa for LaH$_{10}$ synthesis, whereas the second cell (B003) was loaded at 120 GPa to generate the lower stoichiometry superhydride, LaH$_{6}$ (Fig. \ref{figure1.fig}). Both cells were then laser-heated at the HPCAT beamline of the APS. A 20 $\mu$m diameter laser spot was rastered across the sample at fixed positions to promote the dissociation of AB into cBN and the hydrogen, that reacts with the La.  Diffraction patterns were collected at each spot after laser heating. The sample in B002 received about 45 laser pulses (4 to 5 pulses at each of nine points on a 10 x 10$\mu$m grid).  The synthesis for B003 was stopped prematurely after a few pulses for fear that a catastrophic failure of the diamond anvils had occurred (see SI for details). The electrical resistivity was then measured on the samples at NHMFL-Tallahassee, first using a Quantum Design 16 T PPMS. Figure \ref{figure2.fig} shows the first cool down trace from 300 to 230 K for B002 using four of the six available electrodes. A resistance drop is clearly evident at 294 K with no additional transition observed upon cooling to 230 K, the temperature range in which T$_{c}$ has been previously observed for LaH$_{X}$  systems.\\

\begin{figure}[h]
\centering {\includegraphics[scale=0.4]{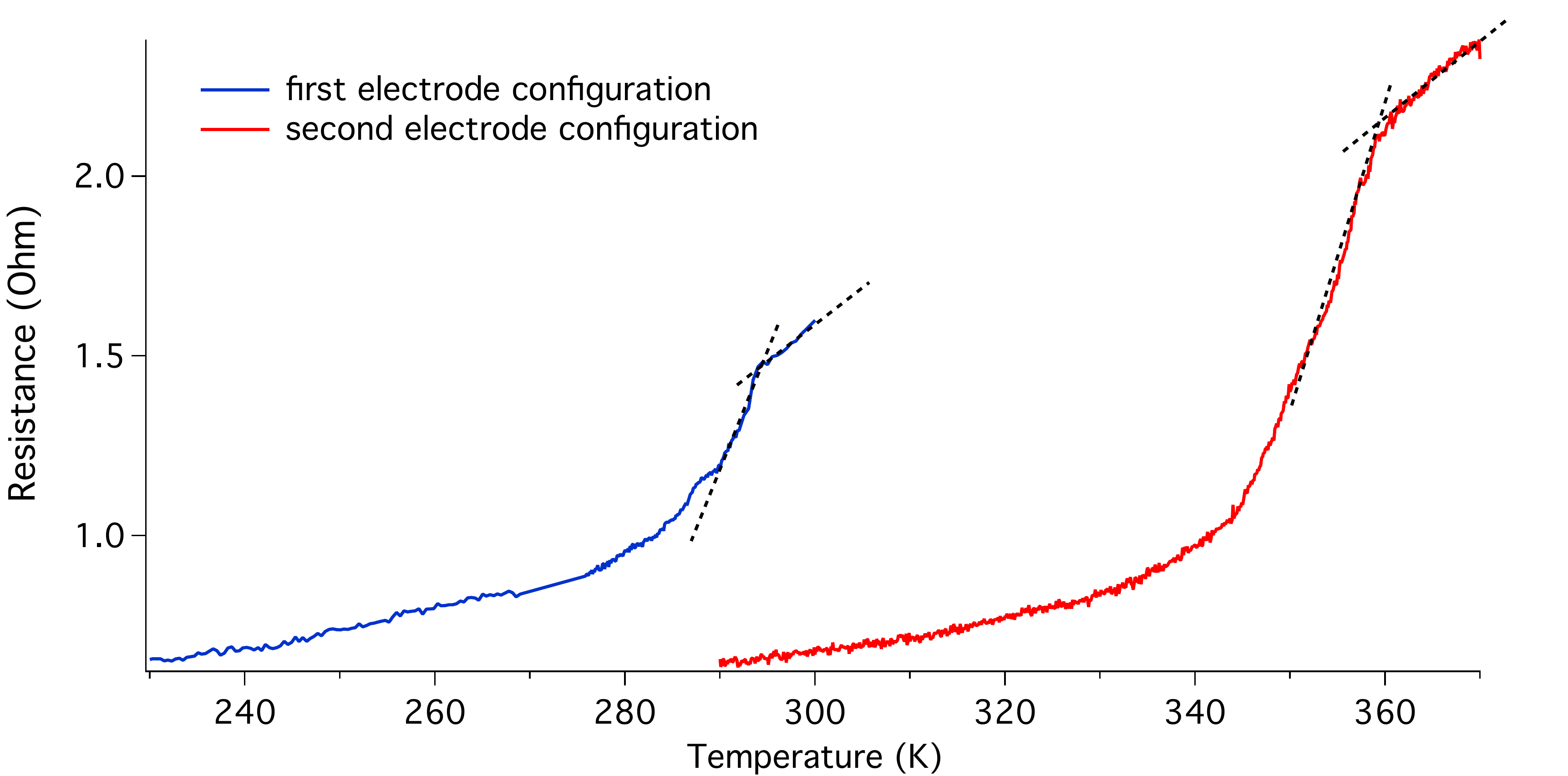}}
\caption{\it  Temperature dependence of resistance R(T) for cell B002 after the initial laser synthesis showing the 294 K temperature transition. The first thermal excursion, done with the intent of observing the normal state more fully, transformed the material and yielded a new onset temperature around 357 K.  This second trace was made using a different lead configuration (see SI). There is a strong background to the signal due to a network of series and parallel contributions.  Measurements on B003, which contained unreacted La, shows the same R(T) dependence and non-zero resistance as low as 1.9 K, which in turn allowed us to subtract out this contribution.  Some portion of the background may also be due to the incomplete nature of the transition and has been seen in other experimental work on LaH systems \cite{Somayazulu2019}$^{,}$\cite{Drozdov2019}. Temperatures indicated are those of the PPMS chamber. }
	\label{figure3.fig}
\end{figure}

In an attempt to measure the resistance of the sample in B002 further into the normal state, we subjected the sample to a series of thermal cycles at 0 T. Surprisingly, these successive higher temperatures excursions pushed the onset of the transition to higher and higher temperatures. We had to stop temporarily at 390 K as this is the maximum temperature possible in our PPMS. Figure \ref{figure3.fig} compares the initial and final traces, with the 0 T onset appearing near 357 K in the final thermal cycle in the PPMS. \\

\begin{figure}[h]
\centering {\includegraphics[scale=0.45]{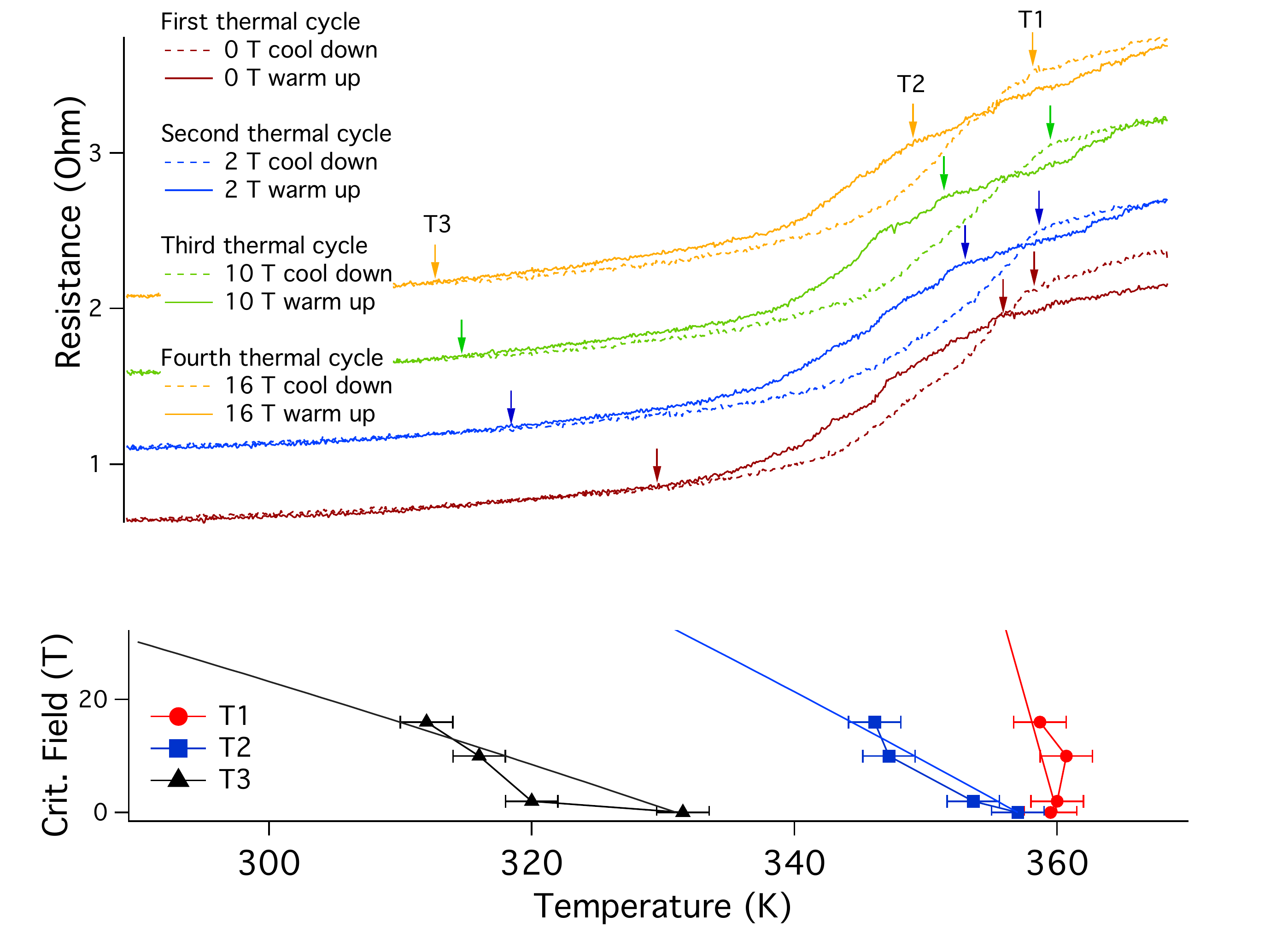}}
\caption{\it \footnotesize a) Top :  temperature dependence of resistance R(T) for B002 on successive cool downs and warm ups at four different fixed applied magnetic fields, 0, 2, 10 and 16 T in a QD PPMS, using a current of 0.6 $\mu$A at 17 Hz between 290 and 370 K at 0.2 K/min with a thermalization time of 30 min at 290 K before warming up. The cool down and warm up traces are indicated by dashed and continuous lines, respectively, and are shifted vertically for clarity. Quantum Design accounts for the magnetoresistance of the Pt thermometer used in the PPMS high temperature option. The arrow on the cool down traces indicate the approximate transition onset and is labeled T1. T2 is the approximate transition onset on the warm up curves, and T3 is the temperature at which the hysteresis of the cool down and warm up traces closes. This run used the 2nd configuration of electrodes (see SI).  The traces were taken sequentially from 0 T to 16 T.  All four traces collapse onto one another below 315 K.  The T$_{c}$ (90/10) is approximately 11 K after a background subtraction is done.  Note that the transition in warming curve for each field is lower in temperature than the cooling.  Although it is possible that this is due to a lag between the thermometer and the sample, the hysteresis is most likely real as the temperature was swept at 0.2 K/min and the separation is different for the four fields. b) Bottom : evolution of T1, T2, and T3, at various magnetic fields, and corresponding fits to the points. }
	\label{figure4.fig}
\end{figure}

Four thermal cycles between 370 K and 290 K were then performed at 0, 2, 10 and 16 T (Fig. \ref{figure4.fig}). 
The traces are shifted vertically for clarity, but collapse onto one another below 310 K. All curves show hysteresis. The warming curve has a lower onset temperature than the cooling curve, indicating either the first order nature of the transition or that the sample is still evolving upon successive thermal cycles. The warm up traces show additional changes which we believe points to a continuing synthesis, as discussed below. \\

To analyze these data, three temperature points are identified in each trace within Fig. 4a. T1 is the transition onset on cool down, T2 is a bump observed at intermediate temperatures on warm up, and T3 is the temperature at which the hysteresis loop closes. A clear shift with applied field is observed for T2 and T3, the former shifting by 11 K and the latter by 19.5 K between 0 and 16 T (Fig. \ref{figure4.fig}a). This shift is evidence of the superconducting nature of this high temperature transition. A fit of the data (fig\ref{figure4.fig}b) using the Ginzburg-Landau relation \cite{Ginzburg1950} (equation 1): 
\begin{equation}
  \mu_{0}H_{c}(T) =  \mu_{0}H_{c}(0)(1 -( \dfrac{T}{T_{c}})^{2})
\end{equation}

yields an  H$_{c}$(0 K)  of 1500 T, 230 T, and 130 T for T1, T2, and T3, respectively. However the error is large for H$_{c}$(0 K) as we had access to a limited field range.  \\

To further investigate the shift in the superconducting transition with field, B002 was then measured in a 41.5 T DC resistive magnet at the NHMFL-Tallahassee. All six working electrodes were used in order to measure an additional resistance channel. We present in the main text only the results from the same configuration as that measured in the PPMS, but the other one yielded similar results (see SI). Successive thermal cycles carried out to yet higher temperatures in order to establish the clear signature of the normal state resulted again in a clear shift of the transition to higher temperature. The stability of epoxies internal to the DAC limited our measurements to below 580 K. \\

\begin{figure}[h]
\centering {\includegraphics[scale=0.45]{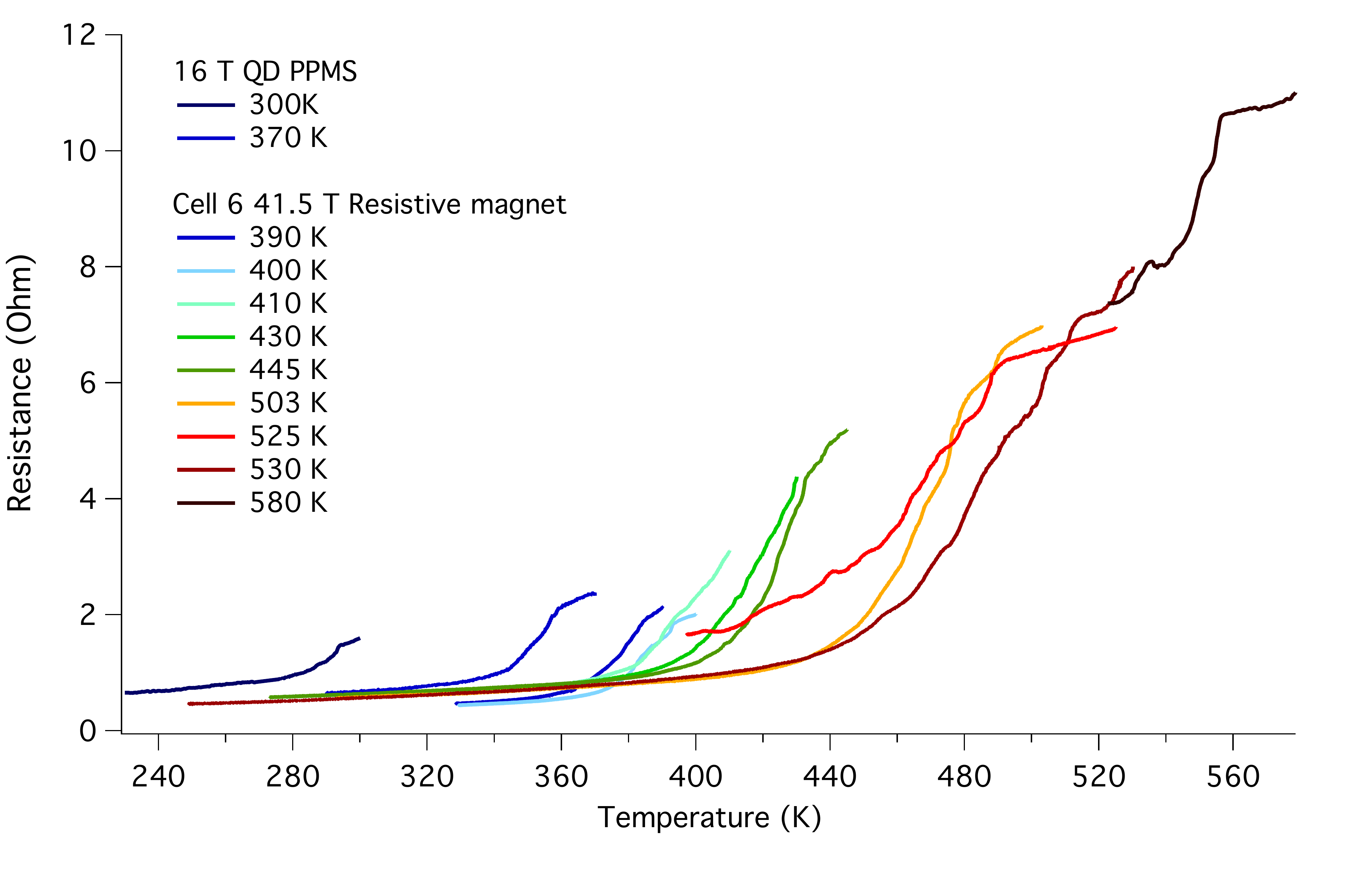}}
\caption{\it  R(T) at 0 T showing the evolution of the superconducting transition with successive thermal cycles (cell B002). These cool down traces were obtained using the same set of electrodes and the same current of 600 nA. The PPMS traces were collected  at 7 Hz, and the cell 6 data at 48.5 Hz to minimize the out-of-phase component. The onset temperature is clearly shifted from 295 K to 560 K, and the transition amplitude also grows with each thermal cycling to higher temperature. The resistance in the proposed superconducting state is also identical for all curves. The temperature excursion to 580 K was realized on April 1 after the end of our magnet time, which was both enabled and stopped by the COVID-19 crisis. }
	\label{figure5.fig}
\end{figure}

Figure \ref{figure5.fig} shows the cool down traces starting from various initial maximum temperatures. The cell remained at each maximum temperature for at least 30 minutes to allow the synthesis to stabilize as realized by a steady sample resistance, and was then cooled at 0.5 K/min. With each higher temperature excursion T$_{c}$ rises further, eventually reaching a maximum value nearing 560 K. The amplitude of the transition also dramatically increases with each excursion to higher temperatures. Similar shifts in transition temperature with repeated thermal cycling has been documented for other superconducting hydrides under pressures \cite{Drozdov2019}$^{,}$\cite{Ahart2017}. The resistance within the superconducting state is almost identical for all the curves, the non-zero state being attributed to the additional unreacted lanthanum between the electrodes, as this same behaviour is seen in B003 (see SI). \\

\begin{figure}[h]
\centering {\includegraphics[scale=0.5]{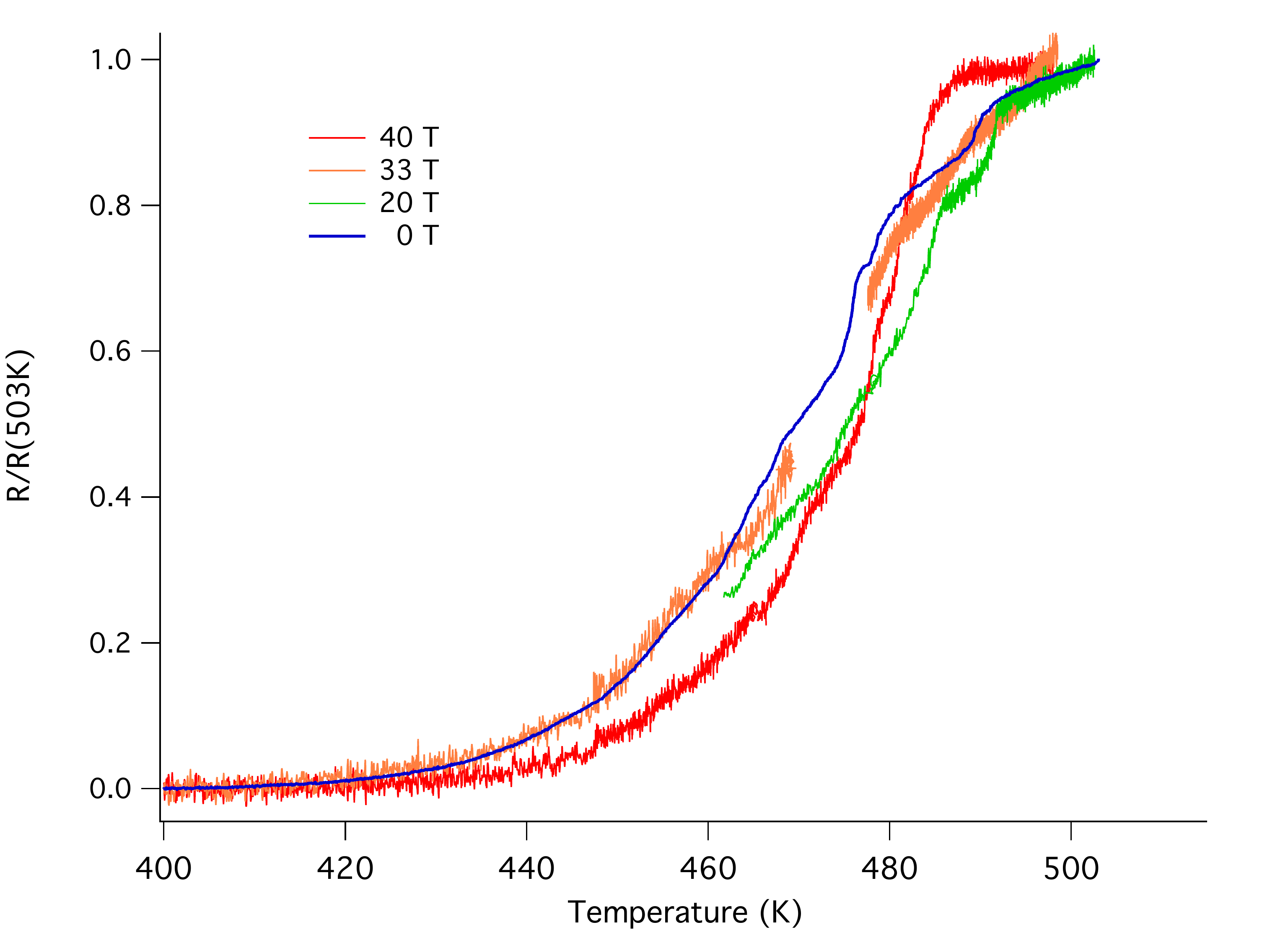}}
\caption{\it  B002 R(T) showing cool down traces in static magnetic fields of 40 T, 33 T,  20 T and 0 T measured in that order. The temperatures in field are corrected for the magnetoresistance of the Pt thermometer used (see SI), and a quadratic background was subtracted. The difference in signal over noise ratio between the zero-field and applied field traces arises from the vibrations generated by the cooling water flowing through the resistive magnet when in operation. The data acquisition software crashed during the 33 T cool down and resulted in an incomplete curve between 470 and 480 K, and the 20 T cool down was shortened by the COVID -19 shutdown.  It is not possible to perform a G-L fit on this data set with any confidence, since the sample continues to evolve with each temperature excursion.}
	\label{figure6.fig}
\end{figure}

Temperature sweeps in static high magnetic fields were performed for initial maximum temperatures of 400 K, 430 K, 445 K, 503 K and 530 K. Only cool down traces were performed in field to remain within the allocated energy budget of a magnet time run. Figure \ref{figure6.fig} shows the result of the run at 503 K, since the data at 530 K indicated either an electrical contact degradation or the development of a touch between the cryostat and the magnet, adding vibrations in the signal which degraded the signal-to-noise. The temperatures for the traces obtained in field are corrected for the magnetoresistance of the thermometer, and a quadratic background in the superconducting state was subtracted over the whole temperature range (see SI). \\

Although the onset of the transition shifts from 492 K at 0 T to 486 K at 40 T, the width of the transition at 0 T, as well as the additional transitions appearing in the 20 T and 33 T curves make the interpretation difficult. The incomplete nature of the chemical transformation precludes fixed-field temperature sweeps within a family of curves above 370 K. Indeed, the shift in the transition with each increasing temperature excursion(Fig. \ref{figure5.fig}) suggests that the chemical state of the material is still evolving.

\section*{Discussion}
We report the observation of superconductivity in a La-based superhydride sample beginning at room temperature that shifts up in a controlled fashion with thermal excursions to a value close to 560 K with a notable concomitant increase in the amplitude. The observation of the onset at 294 K is consistent with preliminary observations of T$_{c}$ above 260 K that were reported previously \cite{Somayazulu2019}. The initial increase in T$_{c}$ from 294 K to 370 K may be related to the enhancement of T$_{c}$ on repeated thermal cycling observed in the simpler binary hydride H$_{3}$Se \cite{Ahart2017} and described in the preliminary reports of the synthesis of superconducting H$_{3}$S \cite{Drozdov2014}. However, this is unlikely, as the maximum predicted transition temperature for the binary LaH$_{10/11}$ system is  288 K \cite{Liu2017}$^{,}$\cite{Peng2017}. Rather, the higher temperature transitions point to additional chemical transformations induced by pressure, shear, temperature, potentially magnetic field, and possibly molten hydrogen, which at room temperature exists above 200 Mbar \cite{Babaev2004}$^{,}$\cite{Liu2018}.  In addition to B and N from the hydrogen source (NH$_{3}$BH$_{3}$) and/or the composite gasket insert (cBN) and the carbon from the epoxy binder, C and Ga from the Pt electrodes also make contact with the La-H and could react with this binary system to form a ternary or higher order system. One might speculate that the initial laser synthesis generated the binary, and the thermal excursions performed on the entire cell may have been critical as this degraded the epoxy binder in the cBN insert, which allowed C and possibly H to diffuse into the sample. Pt and Au from the electrodes and the amorphous FIBed surface of the diamond may also act as catalysts or catalytic bed, respectively, to help form doped alloys or new stoichiometric compounds. A variety of high P-T induced chemical reactions, phase transformations, or novel phases involving these elements are documented, even at more modest conditions \cite{Gregoryanz2004}$^{,}$\cite{Teredesai2004}$^{,}$\cite{Cava1994}. Interestingly, the 294 and 360 K cooldown traces appear to have only two phases with perhaps a third broad phase, which may be due to a disordered phase while the higher onset temperature transitions have numerous transitions in the warming and cooling traces which support the idea of a chemical reaction that is activated with increasing temperatures.  To achieve the higher transition temperatures observed, thermally optimized ordering of the hydrogen in the system might have occurred, possibly with a reduction in the dimensionality or realization of a more crystalline material. Even at 580 K, the reaction is not complete, and higher temperatures and/or further thermal annealing is required.  \\

The high-temperature transition onsets have the characteristic features of superconductivity, but other interpretations cannot be ruled out in the absence of measurements of the Meissner effect, or shifts in T$_{c}$ with magnet field that are at this time complicated by the ongoing chemical transformations taking place during the measurements (e.g., the very high H$_{c}$ implied by the fit to the existing data). Formally, an unusual temperature-induced electronic transition to an insulating state needs to be ruled out, too. However, the transition moves with temperature excursions in an organized fashion from a theoretically predicted superconducting transition of 294 K, and the background and non-zero resistance are convincingly traced back to experimental artifacts. We are left with a material with a zero resistance at very high temperatures. In the absence of additional constraints, our observations provide evidence of  'hot' superconductivity in a still evolving compound (or assemblage), perhaps analogous to recent calculations for the Li-Mg-H system in which T$_{c}$ of 470 K has been predicted within a conventional BCS framework. \\

Additional work is clearly required to characterize the observed phenomena, including structural characterization of the phase (or phases) produced, determination of P-T-H phase diagrams of the phases present, and driving the reaction to completion.  The latter will require some combination of conventional and IR laser heating possibly via fiber optic, which would allow the sample to be imaged and the pressure to be measured at the experimental temperature of interest via Raman. The current experiments were hampered by the unexpected need to explore high temperatures instead of the originally designed range below 300 K, and further characterization was cut short by laboratory closings due to the COVID-19 pandemic. Experiments in higher, pulsed magnetic fields are scheduled to better characterize the field effect on this transition. Beyond the possibility of hot superconductivity, the surprising observations documented here suggest new routes for creating new materials using multiple extreme environments of pressure, temperature and high magnetic fields.

\section*{Methods}
Our piston-cylinder DACs (\diameter 9.5 x 39 mm long, SI Figs. 1 $\&$ 2 (model and photo) are based on concepts presented by Eremets \cite{Eremets1996}. They were made from HIPed NiCrAl to reduce vibrations and eddy current heating due to dB/dt which is on the order of 10 000 to 20 000 T/s in pulsed fields).  72 $\mu$m culet, double bevel (8\textdegree $\times$ 250 $\mu$m, 15\textdegree$\times$ 350 $\mu$m) standard anvils with a 3.75 mm girdle were used.\\
For the preparation of robust conductive leads, we applied a dual beam focused ion beam (FIB) system in combination with a scanning electron microscope (SEM). A metalorganic gas, Trimethyl(methylcylopentadienyl)Platinum(IV), is injected into the high-vacuum sample chamber via the nozzle of a gas injection system (GIS) [SI]. A focused ion stream decomposes the molecules, precisely depositing Pt rich in carbon and gallium \cite{Tao1990} (typically 30 and 10-20 at.$\%$, respectively) onto the anvils. In the same process, the surface of the diamond is amorphized to a depth of approximately 20 nm (for 30 keV gallium ions), allowing the carbon-rich Pt deposit to chemically connect with the broken carbon bonds of the diamond. This chemical bonding results in the mechanical adhesion of FIB-deposits to the diamond, making it extremely robust against mechanical forces. One disadvantage is that the high carbon content reduces the conductivity of such leads significantly (a few Ohms/$\mu$m depending on the deposition conditions and the thickness of the layer).  In order to realize ohmic lead resistances we deposit, in a second step after a transfer to an external sputter deposition system, a layer ($\approx$100 nm) of pure gold on top of the prepared platinum ribbons, with Kapton tape used to protect the diamond surface against Au deposition.  In the third step, we make use of the high-current gallium beam and etch away excess gold until the amorphous diamond surface is recovered between adjacent taps. The platinum-gold ribbons are then covered with an additional FIB-Pt protection layer ($\approx$ 1 $\mu$m) running alongside the pavilion up to the culet.  In a last step thin FIB-Pt ribbons (with approximately 1 $\mu$m thickness) are deposited close to the central part of the culet extending the Pt-Au-Pt “sandwich” leads into the sample space, which has a diameter of approximately 40 $\mu$m.

135 $\mu$m thick aged MP35N gaskets, located on the piston anvil, were indented to a pressure of 15 GPa in a dummy DAC with the same anvil geometry as B002 and B003 to prevent damage to their electrodes yet provide a gasket which mirrored the anvil shape in these DACs.  It was removed and laser cut to a diameter half-way through the second bevel, polished to remove burrs and most of the extruded region, repositioned on the dummy DAC, and then filled with a dry mixture of cBN powder and epoxy (10$\%$ by weight). The gasket was pressed to a load of 25 GPa after which it was laser drilled to a diameter of 40 $\mu$m.  This composite gasket was then moved to the piston of the DAC to be used in the experiment.  The piston with the gasket secured in a crown was brought into a flowing argon filled glove box (O$_{2}$ and H$_{2}$O content of <1 ppm and the gasket hole was filled with ammonia borane (NH$_{3}$BH$_{3}$, or AB) as the hydrogen source and pressure medium. The gasket and gasket crown on the piston are electrically isolated from the rest of the DAC.  Pressure measurements required for these various steps in the gasket fabrication were performed at room temperature using the Raman edge of stressed diamond \cite{Akahama2006}. A third DAC is used to mechanically thin and shear a piece of 99$\%$ La (Goodfellow Metals) to expose clean metal.  A flake of the metal, approximately 4 $\mu$m thick, is extracted from this film and placed on a plastic transfer piston, after which it is brought into the glovebox and pushed against the anvil with FIBed electrodes to form a cold weld.  The same plastic transfer anvil is used to initially align this anvil during the assembly of the DAC and the visible impression that remains of that anvil’s culet helps ensure that the La is positioned over the electrodes in this second step.  To prevent the reaction between La and air, we loaded the sample and sealed the DAC within 30 minutes of extracting the 3-6 $\mu$m samples from the freshly exposed metal.\newline

The assembled DAC was then taken to the desired pressure for synthesis, using the Raman edge of the stressed diamond.  Laser synthesis was performed at HPCAT, sector 16 of the Argonne Photon Source (APS), and XRD analysis of the result was attempted after synthesis.  The sample heating was single-sided using a IPG YLR-100-1064-WC fiber laser operating in modulation mode (square single pulse) with a total power of 100 W. Temperature measurements were obtained from a 4-micron area in the center of the heating spot using an Acton SpectroPro SP2560 imaging spectrograph equipped with a back-illuminated CCD detector (PI-MAX, Princeton Instruments). The laser focal size was 20 $\mu$m$^{2}$ and the modulation pulse width was 30 ms \cite{Meng2015}.
The upstream side of the DAC has a 11.5° degree opening in the piston; the downstream side has a 14° opening in the endcap that supports the anvil with electrodes.  This was not sufficient to allow for confirmation of the binary and the higher order system via X-ray, but a separate experiment is planned for that upon reopening of the APS.  Additional information on sample synthesis and laser heating methods are described in the supplemental information.

AC electrical transport measurements in the PPMS were carried out using the internal electronics (7 Hz) while measurements in the 41.5 T resistive magnet used two Stanford Research Systems SR860 lockin amplifiers in combination with an SRS CS580 voltage controlled current source to drive a 600 nA current at 48.5 Hz.

High magnetic field studies in the 41.5 T resistive magnet at NHMFL-Tallahassee were carried out at fixed fields using consistent sweep rates of 0.5 to 3 K/min in various temperature regions. A custom Janis cryostat with variable temperature insert provided the sample environment.  A small bobbin with three 50 Ohm wire wound heaters in intimate contact with the DAC was mummified in two layers of 12 $\mu$m copper foil and 37 layers of superinsulation.  A Lakeshore Cryotronics Pt-100 thermometer located within the body of the cell and in intimate mechanical contact with the spring and piston was used for both control and sensing.  The maximum temperatures of the measurements were limited as the epoxy used to form electrical connection between the twisted pairs and the FIBed electrodes, EPO-TEK H20E, has a maximum operating temperature of 573 K, began to degrade.  The epoxy used to fix the twisted pairs in the DAC, Stycast 2850 FT with catalyst 24 LV has a maximum operating temperature of 390 K and was completely calcinated after the series of high temperature measurements. 

\bibliographystyle{naturemag}
\bibliography{Superhydride_biblio.bib}

%% Here is the endmatter stuff: Supplementary Info, etc.
%% Use \item's to separate, default label is "Acknowledgements"
\newpage
\begin{addendum}
 \item Portions of this work were performed at HPCAT (Sector 16), Advanced Photon Source (APS), Argonne National Laboratory. HPCAT operations are supported by DOE-NNSA’s Office of Experimental Sciences.  The Advanced Photon Source is a U.S. Department of Energy (DOE) Office of Science User Facility operated for the DOE Office of Science by Argonne National Laboratory under Contract No. DE-AC02-06CH11357. Part of this work was performed at the National High Magnetic Field Laboratory, which is supported by NSF Cooperative Agreement No. DMR-1157490/1644779 and by the State of Florida. M.A., R.K. and R.H. acknowledge funding from the U.S. National Science Foundation (DMR-1933622). We would like to acknowledge stimulating discussions with George Crabtree, Jason Cooley, Pedro Schlottmann, Gil Lonzarich, Thierry Klein, Christophe Marcenat and Rick Wilson. We thank M.König and S.Seifert for their technical support with FIB and sputter deposition. We would like to thank for their help Eric Rod, Curtis Kenney-Benson, Freda Humble, Roald Hoffmann, Andy Rubes, David Sloan, William Brehm, Daniel McIntosh, Alan Williams, Troy Brumm, Bobby Joe Pullum, Scott Maier, Robert Carrier, Christopher Thomas, Michael Hicks, Joel Piotrowski, Larry Gordon, Timothy Murphy and Donna Elliott at SQM. S.W.T., W.A.C. and A.D.G. would like to dedicate this work to the memory of G. Schmiedeshoff.
 \item[Competing Interests] The authors declare that they have no competing financial interests.
 \item[Author contribution statement] S.W.T. and A.D.G. conceived, developed, performed  the experiments and data analysis. Sample preparation was performed by S.W.T., A.D.G., M.A. and R.K. Y.M provided beamline user support. T.H. developed and carried out the FIB process. W.A.C. assisted with the experiment and performed data analysis.  S.W.T., M.O. and V.W. designed and fabricated the DACs. All authors contributed to the discussions and writing. 
\item[Correspondence] Correspondence and requests for materials
should be addressed to A.D.G..~(email: agrockowiak@gmail.com) and S.W.T. ~(email: tozer@magnet.fsu.edu)
\end{addendum}

\newpage

%%
%% TABLES
%%
%% If there are any tables, put them here.
%%

\end{document}